\documentclass{arxiv2}

\begin{document}

\catchline{}{}{}{}{}
%%%%%%%%%%%%%%%%%%%%%%%%%%%%%%%%%%%%%%%%%%%%%%%%%%%%%%%%%%%%%%%%%%%

\title{The mass structure of SU(3) multiplets and pion muon mass difference
}

\author{T. A. Mir$^{*}$ and G. N. Shah }

\address{Nuclear Research Laboratory, Bhabha Atomic Research Centre, \\
Zakura, Srinagar-190 006, Jammu and Kashmir, India\\
$^{*}$taarik.mir@gmail.com}

\maketitle

%\pub{Received (Day Month Year)}{Revised (Day Month Year)}

\section{Abstract}
The mass structure of hadron multiplets is understood to imply the inexactness of SU(3) symmetry. Here we show that these symmetry broken mass splittings amongst baryon and meson multiplet members are close integral multiples of the mass difference between a neutral pion and a muon, the first excitation within the elementary particle mass spectrum. This is found to be equally true for the mass intervals amongst the particles belonging to the multiplets having different spin and parity characteristics. The results reinforce our earlier contention that the mass difference between a neutral pion and a muon is of fundamental importance to the elementary particle mass distribution. 
\section{Keywords}

Hadron multiplets; SU(3); mass splittings.
%\end{abstract}

\ccode{PACS Nos.: 12.10.Kt, 12.40.Yx, 13.40.Dk}

\section{Introduction}	
The SU(3) symmetry and the quark model for the hadrons are the two hallmarks of the standard model. The remarkable order in the properties of hadrons has led to their ogranization into mutilpets according to the so called "Eightfold Way". This classification scheme based on the SU(3) symmetry places the hadrons into families on the basis of spin and parity\cite{Gell-Mann}$^{-}$\cite{Ne'eman}. In fact, explaining the structure of these hadron multiplets led to the proposition of the quark model\cite{Gell-Mann1}$^{-}$\cite{Zweig}. The success of SU(3) symmetry led to the Gell-Mann Okubo\cite{Okubo} and Coleman-Glashow\cite{Coleman} formulae which interelate the masses of pseudoscalar mesons and octet baryons respectively. However, the standard model has neither provided a generally accepted method for calculating the absolute values of strong and electromagnetic mass splittings nor a way of calculating the original masses about which splittings occur\cite{Carter}. This has led to empirical and theoritical investigations based on experimental data  to explore a possible relationship amongst the masses of elementary particle that vary from fractions of electron volts to hundreds of GeVs\cite{Shah}$^{-}$\cite{Akers}. Some of these studies have shown that electron, muon and pion masses serve as basic units for the quantization of mass/mass intervals among elementary particles\cite{Nambu}$^{-}$\cite{Akers}. Further, the elementary particle mass differences have of late been found to have a general tendency to be close integral/half integral multiple of mass difference between a neutral pion and a muon i.e. 29.318 MeV\cite{Shah}. In the present study, we evaluate the applicability of this result to the  baryon and meson multiplets. We reveal that the symmetry broken mass splittings of the hadrons are close integral multiples of the mass difference between a neutral pion and a muon. This reinforces our earlier result that elementary particles do not occur randomly and are linked through the mass difference between first two massive elementary particles i.e. a muon and a neutral pion.
\section{Data Analysis and Results}
The database for the present study is the latest version of the Particle Data Group listings\cite{Yao}. Here we investigate relationship the pion-muon mass difference has with the mass structure of the hadrons which are classified into multiplets on the basis of SU(3) symmetry. According to standard model, hadrons are strongly interacting particles and the lowest-lying hadrons are composits of up, down and strange quarks. The baryons being composed of three quarks whereas mesons are composits of a quark and an antiquark.
\subsection{Baryon Multiplets}
\subsubsection{Baryon Octet}The masses of the baryon octet members with spin J and parity P such that $J^{P}$=$\frac{1}{2}$$^{-}$ are: $m_{p}$=938.27203 MeV, $m_{n}$=939.56536 MeV, $m_{\Lambda^{0}}$=1115.683 MeV, $m_{\Sigma^{+}}$=1189.37 MeV, $m_{\Sigma^{-}}$=1197.449, $m_{\Xi^{0}}$=1314.83 MeV and $m_{\Xi^{-}}$=1321.31 MeV. The successive mass differences are tabulated in Column 1 of Table 1 with numerical value in MeVs given in Column 2. The small mass differences between the different members of an isospin charge multiplet are known to arise from the electromagnetic interaction\cite{Perkins}.
However, the masses of the members of different isospin multiplets differ considerably. Column 4 shows the integral multiples of 29.318 MeV that are close to the observed mass difference between successive members of the octet. The integers being shown in Column 3. The deviations of the observed value from the closest integral multiple of 29.318 MeV are given in Column 5. It is observed that the mass difference between $\Lambda^{0}$ and $n$ i.e. 176.118 MeV differs from the nearest predicted value of 175.908 MeV by only 0.21 MeV. Same is true of the mass difference i.e. 117.381 MeV between the particles $\Xi^{0}$ and $\Sigma^{-}$ which differs from the predicted value of 117.272 MeV by only 0.109 MeV. However, observed mass interval of $\Sigma^{+}$ and $\Lambda^{0}$ differs from the predicted value by about 14.264 MeV. 
Interestingly, this large value turns out to be half integral multiple of the mass difference between a $\pi^{0}$ and a $\mu^{-}$. As can be clearly seen from the row 3 of Table 1, the observed mass difference between $\Sigma^{+}$ and $\Lambda^{0}$ i.e. 73.69 differs from the half integral ($\frac{5}{2}$) multiple of pion and muon mass difference by only 0.39 MeV. The maximum mass splitting within the baryon octet i.e. mass difference of 383.037 MeV, between  the heaviest member $\Xi^{-}$ and the lightest baryon $p$ is close integral multiple of 29.318 MeV, differing from the predicted value by only 1.904 MeV. It may be pointed out that 29.318 MeV multiplicity also holds for the mass intervals among any of the octet members\cite{Shah}. Clearly the 29.318 MeV multiplicity holds with great precision for the baryon octet members.   

\begin{table}[h]
\tbl{The observed baryon octet mass intervals as integral multiple of 29.318 MeV}
{\begin{tabular}{@{}lllllll@{}} \toprule
Particles & Mass Difference & Integer & $N$$\times$29.318  & Obsd - Expd  \\
& (MeV) & $N$  & (MeV) & (MeV)\\ \colrule

$\Lambda^{0}$ - $n$\hphantom{00} & 176.118 & 6 & 175.908 & 0.21 \\ 
\\
$\Sigma^{+}$ - $\Lambda^{0}$\hphantom{00} & 73.687 & 3 & 87.954 & 14.267 \\
& & ($\frac{5}{2}$) & 73.295 & 0.392 \\ 
  & \\ 
$\Xi^{0}$ - $\Sigma^{-}$\hphantom{00} & 117.381 & 4 & 117.272 & 0.109 \\
\\
$\Xi^{-}$ - p\hphantom{00} & 383.038 & 13 & 381.134 & 1.904 \\
\\
\botrule
\end{tabular} \label{ta1}}
\end{table}

\subsubsection{Baryon Decuplet} The analysis for the baryon decuplet members with $J^{P}$=$\frac{3 }{2}$$^{-}$ is detailed in Table 2. It may be pointed out that while all the members of the baryon octet are non-resonant states, for the baryon decuplet all the members execpt for the $\Omega^{-}$ are resonances. Since the Particle Data Group reports an average mass for the four charged states of the $\Delta$ baryons and individual masses for the different charge states of $\Sigma^{*}$ and  $\Xi^{*}$ baryons, we cosider the average masses of each isospin multiplet which are as follows: $m_{\Delta}$=1232 MeV, $m_{\Sigma^{*}}$=1384.56 MeV,  $m_{\Xi^{*}}$=1533.4 MeV, and $m_{\Omega^{-}}$=1672.45 MeV.
The first three mass differences in the Table 2 are those between the successive members of the decuplet. The equal spacing rule for the SU(3) decuplet predicts masses of successive isospin multiplets to be equidistant\cite{Oh} i.e. $m_{\Sigma^{*}}$ - $m_{\Delta}$ = $m_{\Xi^{*}}$ - $m_{\Sigma^{*}}$ = $m_{\Omega^{-}}$ - $m_{\Xi^{*}}$. However, this rule is not strictly obeyed in the decuplet, since the mass separations are not exactly same as evident from Table 2. Further, although the mass spacing among the successive decuplet members deviate from the closest integral (5) multiples of 29.318 MeV by 5.97, 2.25 and 7.54 MeV respectively, it is important to note that the average mass spacing among successive members i.e. 146.816 MeV is very close to 146.59 MeV, a value obtained on integral (5) multiplication of 29.318 MeV. The difference between the observed and predicted values being 0.226 MeV only. This may be compared with 140 MeV pionic mass interval\cite{Akers} among the decuplet members which deviates from the average mass spacing by about 6.816 MeV\cite{Mac Gregor4}. The $\Omega^{-}$ - $\Delta$ is the  difference between the mass of lightest resonance member $\Delta$ and the heaviest non-resonant member $\Omega^{-}$ of the decuplet. This observed mass interval of 440.45 MeV is very close to 439.77 MeV, obtained on integral (15) multiplication of the mass difference between a neutral pion and a muon. The difference between the observed and expected value being only 0.68 MeV. 
\begin{table}[h]
\tbl{The observed baryon decuplet mass intervals as integral multiple of 29.318 MeV}
{\begin{tabular}{@{}lllllll@{}} \toprule
Particles & Mass Difference & Integer & $N$$\times$29.318 & Obsd - Expd \\
& (MeV) &  $N$  & (MeV) & (MeV) \\ \colrule

$\Sigma^{*}$ - $\Delta$ \hphantom{00} & 152.56 & 5 & 146.59 & 5.97 \\ 
\\
$\Xi^{*}$ - $\Sigma^{*}$\hphantom{00} & 148.84 & 5 & 146.59 & 2.25 \\
\\ 
$\Omega^{-}$ - $\Xi^{*}$\hphantom{00} & 139.05 & 5 & 146.59 & 7.54 \\
 \\
Average decuplet mass spacing\hphantom{00} & 146.816 & 5 & 146.59 & 0.226\\
\\
$\Omega^{-}$ - $\Delta$\hphantom{00} & 440.45 & 15 & 439.77 & 0.68 \\
\\

\botrule

\end{tabular} \label{ta1}}
\end{table}

\subsection{Meson Multiplets}
\subsubsection{Pseudoscalar Meson Nonet} The mesons with spin zero and odd parity i.e. $J^{P}$ = $0^{-}$ are organized into a multiplet contaning nine states to form pseudoscalar meson nonet. These mesons have the lowest rest energy. In column 1 of Table 3, the k$^{\pm}$ - $\pi^{\pm}$, $\eta$ - k$^{\pm}$ and $\eta^{'}$ - $\eta$ are the mass difference between the successive members of the pseudoscalar meson nonet with numerical values  given in column 2. The deviation of 0.182 MeV between the observed and predicted $\eta^{'}$ - $\eta$ mass difference may be compared with that of 0.730 MeV obtained from the integral multiple (3) of the 137 MeV\cite{Mac Gregor4}. Both the observed mass intervals of $\eta$ - $\pi^{0}$ and $\eta$ - $\pi^{\pm}$ deviate from the 14 multiples of 29.318 MeV by about 2 MeV. However the difference 410.2364 MeV between the mass of $\eta$ and average pion mass of 137.27339 MeV differs by only 0.215 MeV from predicted value 410.452 obtained by the multiplication of 29.318 MeV by integer 14. Again this departure of the observed mass difference between $\eta$ and average pion mass may be comapared with the difference of 0.736 MeV bewteen the observed and that expected when mass intervals are taken as integral multiples of average pion mass i.e. 137 MeV\cite{Mac Gregor4}.  The observed mass interval between the $\eta^{'}$ and average pion mass is 820.506 MeV. This value deviates from the predicted value of 820.904 MeV obtained from the integral multiple (28) of 29.318 MeV by only 0.3973 MeV only. Whereas the predicted value 822 MeV on the basis of integral pion mass differs from the observed value by 1.4933 MeV\cite{Mac Gregor2}. The difference $\eta^{'}$ - $\pi^{0}$ is the mass difference between the lightest and heaviest member of the pseudoscalar meson nonet. As can be seen from the Table 3, the observed mass spacings are close integral multiples of the mass difference between a neutral pion and a muon. The masses for the pseudoscalar mesons taken from the Particle Data Group listings are: $m_{\pi^{0}}$=134.9766 MeV, $m_{\pi^{\pm}}$=139.57018 MeV, $m_{k^{\pm}}$=493.677 MeV, $m_{k^{0}}$= 497.648 MeV, $m_{\eta}$=547.51 MeV, $m_{\eta^{'}}$=957.78 MeV.
\begin{table}[h]
\tbl{The observed pseudoscalr meson mass intervals as integral multiple of 29.318 MeV}
{\begin{tabular}{@{}lllllll@{}} \toprule
Particles & Mass Difference & integer  & $N$$\times$29.318  & Obsd - Expd  \\
& (MeV) & $N$ &  (MeV) & (MeV)\\ \colrule
k$^{\pm}$ - $\pi^{\pm}$\hphantom{00} & 354.107 & 12 & 351.816 & 2.291 \\
\\

$\eta$ - k$^{\pm}$\hphantom{00}   & 53.833   & 2 & 58.636 & 4.803   \\
\\
$\eta^{'}$ - $\eta$\hphantom{00} &    410.27    & 14 & 410.452 & 0.182 \\
\\
$\eta$ - $\pi^{0}$\hphantom{00} &    412.5344    & 14 & 410.452 & 2.0814 \\ 
\\
$\eta$ - $\pi^{\pm}$\hphantom{00} &  407.93982    & 14 & 410.452 & 2.51218 \\ 
\\
$\eta$ - $\pi_{avg}$\hphantom{00} &  410.2364    & 14 & 410.452 & 0.215 \\ 
\\
$\eta^{'}$ - k$^{\pm}$\hphantom{00} &  464.103    & 16 & 469.088 & 4.985 \\
\\
$\eta^{'}$ - $\pi^{0}$\hphantom{00} & 822.81 & 28 & 820.904 & 1.906 \\
\\
$\eta^{'}$ - $\pi^{\pm}$\hphantom{00} & 818.20982 & 28 & 820.904 & 2.69418 \\
\\
$\eta^{'}$ - $\pi_{avg}$\hphantom{00} & 820.5067 & 28 & 820.904 & 0.3973 \\
\\ \botrule

\end{tabular} \label{ta1}}
\end{table}
\subsubsection{Vector Meson Nonet}The nine vector mesons with spin one and odd parity i.e. $J^{P}$ = $1^{-}$ form a multiplet called vector meson nonet. The analysis for the mass differences among the vector meson nonet members is detailed in Table 4. The mass difference between successive members (Column 1) of the vector meson nonet  is given in the Column 2. As is evident from the Table 4, the observed mass differences k$^{*}$ - $\rho$ and $\phi$ - $\omega$ are close integral multiples of 29.318 MeV. Although the mass difference between isospin triplet $\rho$ meson and isospin singlet $\omega$ meson 7.15 MeV is nonelectromagnetic in origin but is of the order of electromagnetic mass splitting. The deviation of 2.226 MeV of the observed $\phi$ - $\omega$ mass difference from the integral multiple of 29.318 MeV is in better agreement than that of 8.19 MeV obtained when mass interval is predicted as half integral multiple of a mass unit of about 70 MeV i.e about half the pion mass\cite{Mac Gregor2}. Since the Particle Data Group lists average mass for the isospin triplet $\rho$ states and separate masses for two charge states of k$^{*}$ meson, we consider the average mass of vector mesons for the analysis which are: $m_{k^{*}}$=893.83 MeV, $m_{\rho}$=775.5 MeV, $m_{\omega}$=782.65 MeV and $m_{\phi}$=1019.460 MeV.

\begin{table}[h]
\tbl{The observed vector meson mass intervals as integral multiple of 29.318 MeV}
{\begin{tabular}{@{}lllllll@{}} \toprule
Particles & Mass Difference & integer  & $N$$\times$29.318  & Obsd - Expd  \\
& (MeV) & $N$ &  (MeV) &  (MeV) \\ \colrule
k$^{*}$ - $\rho$\hphantom{00} & 118.33 & 4 & 117.272 & 1.058 \\
\\

$\omega$ - $\rho$\hphantom{00}   & 7.15   &  &  &   \\
\\
$\phi$ - $\omega$\hphantom{00} &    236.81    & 8 & 234.544 & 2.266 \\

\\ \botrule

\end{tabular} \label{ta1}}
\end{table}
\subsection{Inter multiplet mass intervals}
The applicabilty of 29.318 MeV multiplicity of elementary particle mass intervals extends beyond the mass differences between the successive members of a particular SU(3) multiplet and applies to mass intervals among the members of multiplets with different spin and parity characteristics. The analysis for the mass intervals between the octet and decuplet baryons is detailed in Table 5. The lightest member of decuplet i.e $\Delta$(1232) resonance is the first excited state of the proton and the observed mass interval between the two is accounted by the hyperfine splitting due to colour magnetic interaction among the quarks\cite{Perkins}. However, from Table 5 it is seen that the observed mass interval of 293.728 MeV between $\Delta$(1232) and proton is very close to 293.180 MeV, a value obtained on integral multiplication of 29.318 by 10. The difference between the two values being only 0.548 MeV. The relation  of $\Delta$(1232) and proton is expected as $\Delta$ baryons are considered to be the excited states of nucleon but there is no relation between $\Delta$(1232) and $\Lambda^{0}$. However, from our analysis it follows that the observed mass interval between the two 116.317 MeV differs from 117.272 MeV, the closest integral (4) multiple of 29.318 MeV by 0.955 MeV only. Thus $\Delta$(1232) can be obtained by taking four excitations of 29.318 MeV from the $\Lambda^{0}$. Similarly the observed mass inteval of 556.767 bewteen two unrelated baryons $\Omega^{-}$ and $\Lambda^{0}$ deviates from 557.042 MeV, a value obtained as 19 multiples of pion-muon mass difference by only 0.275 MeV. Further, the observed mass difference 351.14 MeV between the heaviest member of baryon octet $\Xi^{-}$ and that of baryon decuplet i.e. $\Omega^{-}$ the only non-resonant member, differs from 351.816 MeV, a value obtained on integral (12) multiplication of 29.318 MeV, by only 0.676 MeV. The detailed anlysis of the mass differences among the pseudoscalar and vector mesons is given in Table 6. As seen from the Table 6 the 29.318 MeV mass interval multiplicity is valid for mesons also. From Tables 5 and 6 it is clear that the hadron mass intervals fall into two classes 1) those integral multiples of the mass difference between a neutral pion and muon and 2) those in which the difference between the observed and predicted values are large but turn out to be exact half integral multiples of 29.318 MeV. 
\begin{table}[h]
\tbl{The observed mass intervals between octet and decuplet barayons as integral multiple of 29.318 MeV}
{\begin{tabular}{@{}lllllll@{}} \toprule
Particles & Mass Difference & Integer & $N$$\times$29.318 & Obsd - Expd\\
& (MeV) & $N$  & (MeV) & (MeV)\\ \colrule

$\Delta$(1232) - p\hphantom{00} & 293.728 & 10 & 293.180 & 0.548 \\ 
\\

$\Delta$(1232) - $\Lambda^{0}$\hphantom{00} & 116.317 & 4 & 117.272 & 0.955 \\ 
\\
$\Delta$(1232) - $\Sigma^{-}$\hphantom{00} & 34.551 & 1 & 29.318 & 5.233 \\ 
\\
$\Delta$(1232) - $\Xi^{-}$\hphantom{00}& 89.31 & 3 & 87.954 & 1.356 \\
\\
$\Sigma^{*+}$(1382.8) - n\hphantom{00} & 443.15 & 15 & 439.77 & 3.38 \\
\\
$\Sigma^{*+}$(1382.8) - $\Lambda^{0}$\hphantom{00} & 267.117 & 9 & 263.862 & 3.255 \\
\\
$\Sigma^{*+}$(1382.8) - $\Sigma^{0}$\hphantom{00} & 190.158 & 7 & 205.226 & 15.068 \\
& & ($\frac{13}{2}$) & 190.567 & 0.409 \\ 
  & \\ 
$\Sigma^{*}_{avg}$(1384.56) - $\Sigma_{avg}$(1193.153)\hphantom{00} & 191.407 & 7 & 205.226 & 13.816 \\
& & ($\frac{13}{2}$) & 190.567 & 0.840 \\ 
\\
$\Sigma^{*+}$(1382.8) - $\Xi^{-}$\hphantom{00} & 61.49 & 2 & 58.636 & 2.854\\
\\
$\Sigma^{*-}$(1387.2) - $\Xi^{0}$\hphantom{00} & 72.37 & 2 & 58.636 & 13.734\\
& & ($\frac{5}{2}$) & 73.259 & 0.925 \\
\\
$\Xi^{*0}$(1531.8) - n\hphantom{00} & 592.234 & 20 & 586.36 & 5.87\\
\\
$\Xi^{*-}$(1535) - $\Sigma^{-}$\hphantom{00} & 337.551 & 12 & 351.816 & 14.265\\
& & ($\frac{23}{2}$) & 337.157 & 0.394 \\
\\ 
$\Xi^{*-}$(1535) - $\Xi^{0}$\hphantom{00} & 220.17 & 7 & 205.226 & 14.944\\
& & ($\frac{15}{2}$) & 219.885 & 0.285 \\
\\ 
$\Omega^{-}$ - n\hphantom{00}& 733.884 & 25 & 732.95 & 0.934 \\
\\
$\Omega^{-}$ - $\Lambda^{0}$\hphantom{00}& 556.767 & 19 & 557.042 & 0.275 \\
\\
$\Omega^{-}$ - $\Xi^{-}$\hphantom{00}& 351.14 & 12 & 351.816 & 0.676 \\
\\
\botrule

\end{tabular} \label{ta1}}
\end{table} 
\begin{table}[h]
\tbl{The observed mass intervals between pseudoscalar and vector mesons as integral multiple of 29.318 MeV}
{\begin{tabular}{@{}lllllll@{}} \toprule
Particles & Mass Difference & integer  & $N$$\times$29.318  & Obsd - Expd  \\
& (MeV) & $N$ &  (MeV) & (MeV)\\ \colrule
k$^{*0}$(896) - $\pi^{0}$\hphantom{00} & 761.03 & 26 & 762.263 & 1.233 \\
\\
k$^{*\pm}$(891.66) - k$^{\pm}$\hphantom{00} & 397.983 & 14 & 410.452 & 12.469 \\
& & ($\frac{27}{2}$) & 395.793 & 2.19
\\
\\
k$^{*0}$(896) - $\eta$\hphantom{00} & 348.49 & 12 & 351.816 & 3.326 \\
\\
k$^{*0}$(896) - $\eta^{'}$\hphantom{00} & 61.73 & 2 & 58.636 & 3.094 \\
\\
$\rho$ - k$^{0}$\hphantom{00} & 277.852 & 10 & 293.18 & 15.328 \\
& & ($\frac{19}{2}$) & 278.521 & 0.669
\\
\\
$\omega$ - $\pi^{0}$\hphantom{00}   & 647.6734   & 22 & 644.996 & 2.677   \\
\\
$\omega$ - k$^{\pm}$\hphantom{00} &   288.823    & 10 & 293.18 & 4.207 \\
\\
$\omega$ - $\eta$\hphantom{00} &    235.14    & 8 & 234.544 & 0.596 \\ 
\\
$\omega$ - $\eta^{'}$\hphantom{00} &  175.13    & 6 & 175.908 & 0.778 \\ 
\\
$\phi$ - $\pi^{\pm}$\hphantom{00} &  879.89   & 30 & 879.54 & 0.350 \\ 
\\
$\phi$ - k$^{\pm}$\hphantom{00} &  525.783    & 18 & 525.724 & 1.94 \\
\\
$\phi$ - $\eta$\hphantom{00} & 471.95 & 16 & 469.088 & 2.862 \\
\\
$\phi$ - $\eta^{'}$\hphantom{00} & 61.68 & 2 & 58.636 & 3.044 \\
\\
 \botrule

\end{tabular} \label{ta1}}
\end{table}
\section{Discussion}
The assumption that a small set of particles representing the lowest level of elementary particle mass spectrum are the ground states and higher mass elementary particles are their excitations can be found in many studies\cite {Mac Gregor4}$^{,}$ \cite{Rodrigues}$^{,}$ \cite{Barut}. The neutral pion and muon are the two lightest particles with life time more than any other massive particles except for electron and nucleons. The presence of pions and muons in the decay products of most of the elementary particles points to some fundamental relation of these particles to all other higher elementary particle mass states\cite{Mac Gregor3} and studies have been dedicated to pin down this fundamental relation\cite{Nambu}$^{-}$\cite{Akers}. The mass intervals among elementary particles have been shown to be integral multiples of the pion mass\cite {Mac Gregor2} and this integral multiplicity is also true for masses which are reported to be integral/half integral multiples of the pion mass \cite{Tangherlini}$^{-}$\cite{Koschmieder}. Similarly various sequances of particles are reported with mass differences as multiples of the muon mass\cite{Sternheimer1}. The mass unit of 35 MeV equal to one-third of muon mass and one-fourth of the pion mass serves as basic unit for the quantization of the elementary particle masses\cite{Palazzi}. On the other hand we reveal the relevance of mass difference between a neutral pion and a muon to the elementary particle mass spectrum. The mass interval between neutral pion (a hadron) and muon (a lepton) is the first excitation in elementary particle mass spectrum and we show that inter particle mass excitations are related to this basic mass just like the higher excitation energies in atomic levels are related to the first excitation.\\
According to the formula E = mc$^{2}$, c being the velocity of light, to rest mass m of any particle corresponds the stationary level of energy E and discreteness of spectrum of masses of elementary particles is comparable to a discrete spectrum of excitation energies of atom\cite{Nambu}$^{,}$ \cite{Koschmieder}$^{,}$ \cite{Kyriakos}. Important features of mass distribution of elementary particles are revealed on applying the results of the splitting of the atomic levels to the splitting of the mass among the elementary particles\cite{Akers}$^{,}$ \cite{Perkins} although the reasons and calculation of the splitting of the atomic levels are well understood whereas a generally accepted formalism for calculating the absolute values of strong and electromagnetic mass splittings is absent\cite{Carter}.\\
We have satisfactorily shown that the mass structure among the elementary particles, irrespective of their classification, can be explained in terms of the mass difference between a neutral pion and a muon. The mass differences among the elementary particles in general when arranged in the ascending order of mass have a striking tendency to be integral/half integral multiples of 29.318 MeV\cite{Shah}. Same is shown to be true for the mass differences between any of the baryons and mass interval between the unstable charged leptons. From the present study, it follows that the "fine structure" of mass i.e. mass differences betweend different isospin members of a given hadron multiplet attributed to SU(3) symmetry breaking are close integral multiples of the mass difference between a neutral pion and a muon. This is also true for the maximum mass splittings within each multiplet i.e. the mass separation of the lightest and heaviest member. The mass intervals between the members of different SU(3) multiplets, i.e. between octet and decuplet baryons and between the pseudoscalar and vector mesons, believed to be due to the spin-spin interaction\cite{Green}, are also integral multiples of the mass unit of 29.318 MeV. The 29.318 MeV multiplicity of the elementary particle mass intervals clearly indicates that the mass difference between a neutral pion (hadron) and a muon (lepton) is a fundamental mass unit for inter particle mass excitations among leptons, mesons and baryons.
The excellent agreement of the observed elementary particle mass intervals and those calculated as integral multiples of the neutral pion and muon mass difference seems to be not a mere coincidence and is not explained by current models. Clearly the occurence of elementary particle states is not random but seems to follow a definite order such that the mass excitations from one particle to other are always in units of 29.318 MeV i.e. the mass difference between a neutral pion and a muon. 
\section{Conclusion}
The SU(3) symmetry breaking mass splitting among the baryon octet members have been found to be close integral multiples of the mass difference between a neutral pion and a muon. This also holds true for the average mass separation among the successive members of the baryon decuplet, for the mass intervals among the pseudoscalar meson nonet and among the vector meson nonet members. The phenomenon  responsible for the mass structure of hadrons does not effect the validity of the integral multiplicity.  We have revealed that the mass unit of 29.318 MeV i.e mass difference between a neutral pion and a muon is basic to the mass structure of SU(3) hadron multiplets as well. Thus pion and muon mass difference appears to be a universal and a fundamental mass quantum for the elementary particle mass spectrum.

%\section*{Acknowledgments}
%The authors are highly thankful to Paolo Palazzi, Amjad Hussain Shah Gilani and David Akers for their helpful comments and suggestions on the contents of the paper. 

%This section should come before the References. Dedications and funding 

%\section*{References}

\end{document}